\documentclass{PoS}
\usepackage{amsmath}
\title{Towards a determination of the low $x$ gluon via exclusive $J/\psi$ production}

\ShortTitle{Towards a determination of the low x gluon via exclusive $J/\psi$ production}

\author{\speaker{C.~A.~Flett}$^a$, S.~P.~Jones$^b$,  A.~D.~Martin$^c$, M.~G.~Ryskin$^{c,d}$ and T.~Teubner$^a$\\
\llap{$^a$} Department of Mathematical Sciences, University of Liverpool, Liverpool L69 3BX, U.K.\\
\llap{$^b$} Theoretical Physics Department, CERN, Geneva, Switzerland \\
\llap{$^c$} Institute for Particle Physics Phenomenology, Durham University, Durham DH1 3LE, U.K.\\
\llap{$^d$} Petersburg Nuclear Physics Institute, NRC Kurchatov Institute, Gatchina, St. Petersburg, 188300, Russia \\
E-mail: \email{c.a.flett@liverpool.ac.uk}, \email{s.jones@cern.ch}, \email{a.d.martin@durham.ac.uk}, \email{ryskin@thd.pnpi.spb.ru}, \email{thomas.teubner@liverpool.ac.uk}
}

\abstract{We discuss how the stability of the theoretical prediction for exclusive $J/\psi$ photoproduction has been improved through a systematic taming of the known $\overline{\text{MS}}$ coefficient functions by accounting for a formally power suppressed, but numerically significant, correction
encoded within a $Q_0$ cut. The phenomenological implications of this will be emphasised meaning, ultimately, the possibility to include the exclusive data into a global fitter framework to provide constraints on the small $x$ gluon.}

\FullConference{XXVII International Workshop on Deep-Inelastic Scattering and Related Subjects - DIS2019\\
		8-12 April, 2019\\
		Torino, Italy}

\begin{document}

\section{Introduction}

It is well known that the global parton distribution function (PDF) analyses do not well constrain the low $x$ gluon distribution - for $x$ typically below the region accessible to HERA, the global fits are plagued by large uncertainties that are consistent with unphysical decreasing gluon densities. 
We study exclusive $J/\psi$ photoproduction, as measured at HERA and the LHCb via ultraperipheral $p + p \rightarrow p + J/\psi + p$ events, as a means of exploring and ultimately providing constraints in this highly unconstrained low $x$ regime.  

In Section 2, we first illustrate our model framework and briefly recall the challenges one faces in including such data into global fits. In Section 3, we outline how a systematic taming of the naive $\overline{\text{MS}}$ next-to-leading-order (NLO) calculation within collinear factorisation helps to improve the reliability and stability of the theoretical result. Broadly, the implementation of a power correction via a $Q_0$ cut and a scale fixing procedure allow for elimination of crucial double counting effects alongside resummation of a class of large logarithmic contributions which collectively suppress the wild scale variation as seen in the pure $\overline{\text{MS}}$ approach. We further 
demonstrate cross section stability with respect to scale variations and convey, in closing, an indication about the behaviour of gluon densities extracted in global fit analyses within the small $x$ domain. 

\section{Model framework and challenges}
The utility of $p + p \rightarrow p + J/\psi + p$ as a probe of the low $x$ domain can be traced back to \cite{Ryskin}. There the exclusive cross section for the hard quasi-elastic subprocess $\gamma^* p \rightarrow J/\psi p$, which drives the $pp$ initiated reaction, was derived in the leading log approximation (LLA) of perturbative QCD (pQCD), showcasing the dependence of the process on the {\it square} of the gluon distribution.

The cross section for ultraperipheral production, $p+p \rightarrow p + J/\psi + p$, is modelled as \begin{equation}\frac{\text{d} \sigma^{\text{th}}(pp)}{\text{d}y} = S^2(W_+) N_+ \sigma^{\text{th}}_+ (\gamma p) + S^2(W_-) N_- \sigma^{\text{th}}_- (\gamma p),\end{equation}with $S^2(W_{\pm})$ and $N_{\pm}$ survival factors and photon fluxes, respectively, for $\gamma p$ centre of mass energies $W_{\pm}.$ The $\sigma^{\text{th}}_{\pm} (\gamma p)$ are the corresponding photoproduction subprocess cross sections. The two contributions arise from the lack of forward proton tagging where we do not know which initial state proton the photon initiating the (semi) hard process came from, see Fig. 1. 
\begin{figure}[h]
    \centering
    \includegraphics[scale=0.53]{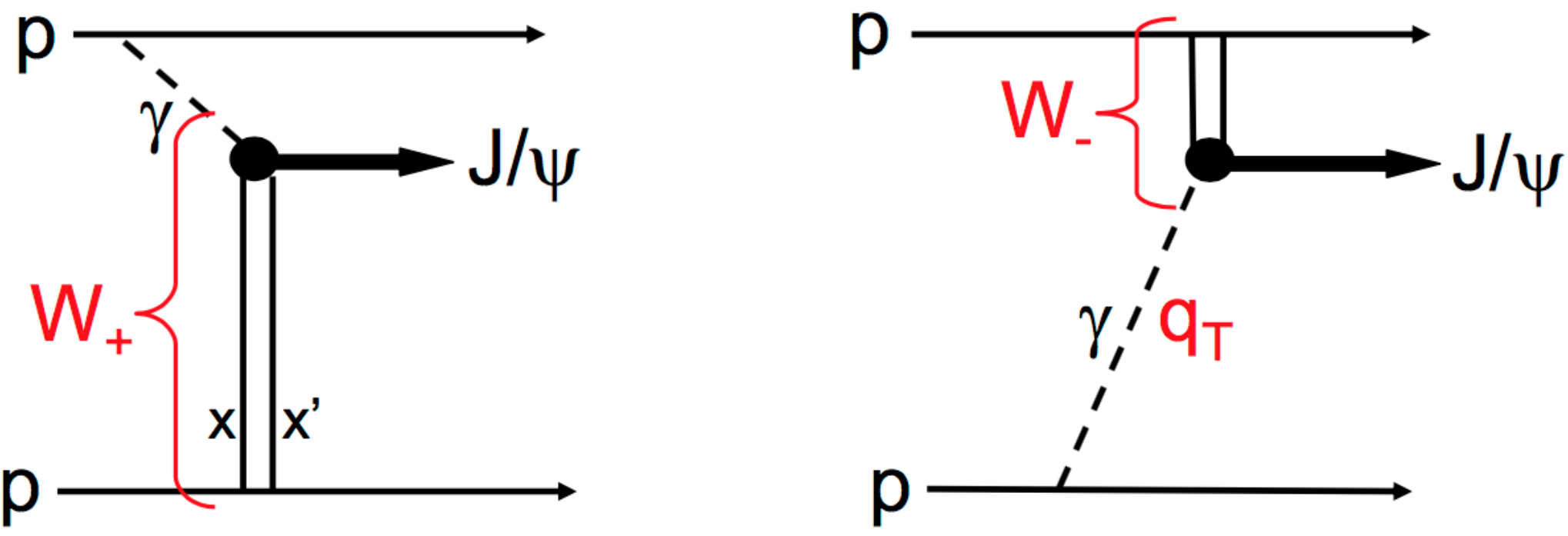}
    \caption{The two independent subprocesses contributing to the ultraperipheral cross section. The exchange at energy $W_+$ allows for a probe of smaller $x$, while the $W_-$ exchange is typically at larger $x$, where there is already significant data support from HERA. The $q_T$ of the photon is small.}
    \label{fig:my_label}
\end{figure}
As $W^2_{\pm} = M_{J/\psi} \sqrt{s} e^{\pm |y|}$, the recent LHCb data at $\sqrt{s} = 13\, \text{TeV}$ \cite{Ronan} allow sampling in the region $x \sim 10^{-5}$ and $x \sim 10^{-2}$ for the $W_+$ and $W_-$ components respectively, with $M_{J/\psi}$ the mass and $y \sim 4$ the rapidity of the $J/\psi$. We can therefore fix the $W_-$ component from a fit to HERA data and extract the $W_+$ component of the cross section using the LHCb data provided in differential bins with respect to $y$, i.e the l.h.s of (2.1).

Our set up for the underlying (semi) hard scattering follows \cite{Ivanov} \cite{dlog} and is given in Fig. 2. 
\begin{figure}[h]
    \centering
    \includegraphics[scale=0.47]{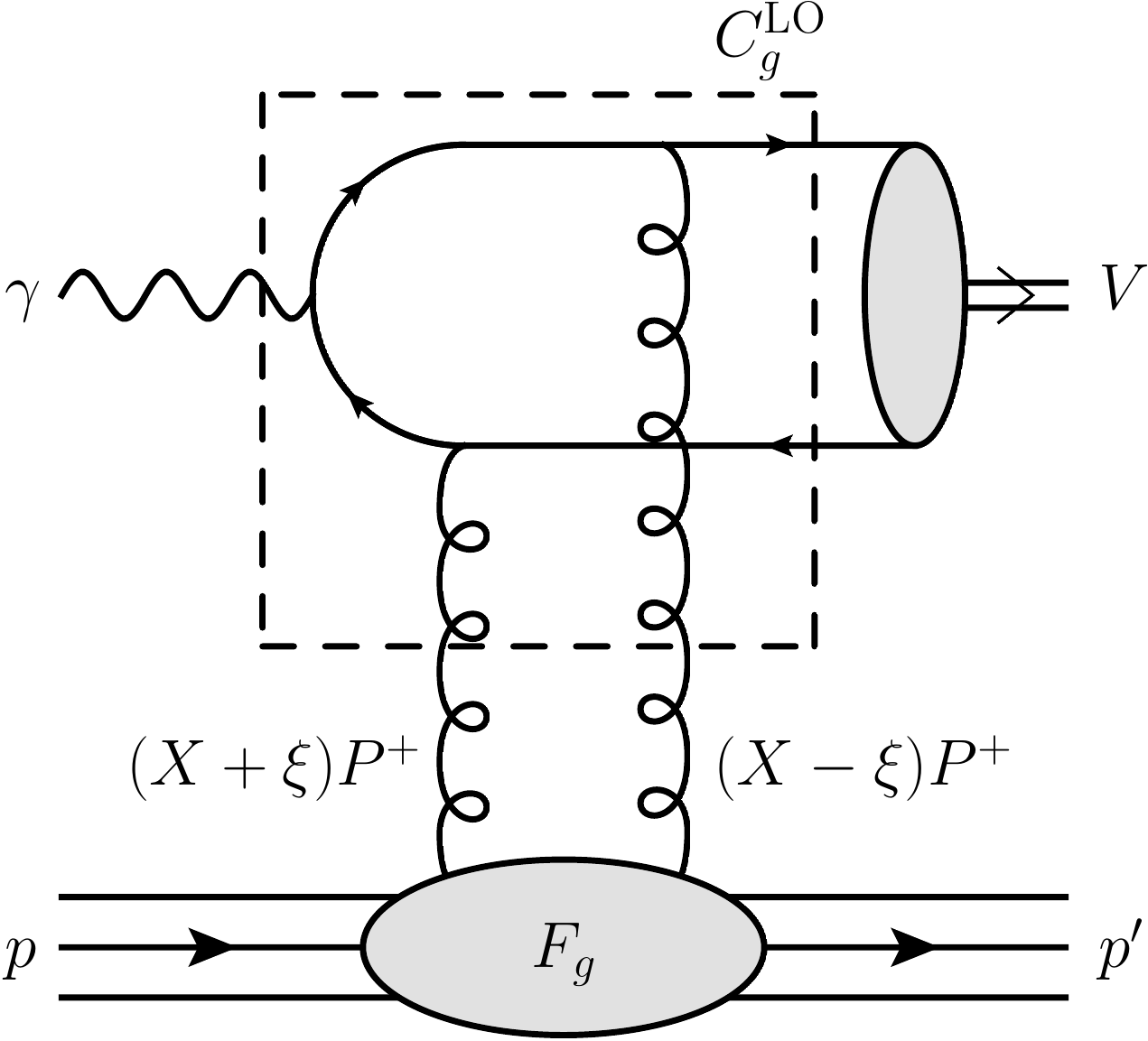}
    \quad
    \includegraphics[scale=0.47]{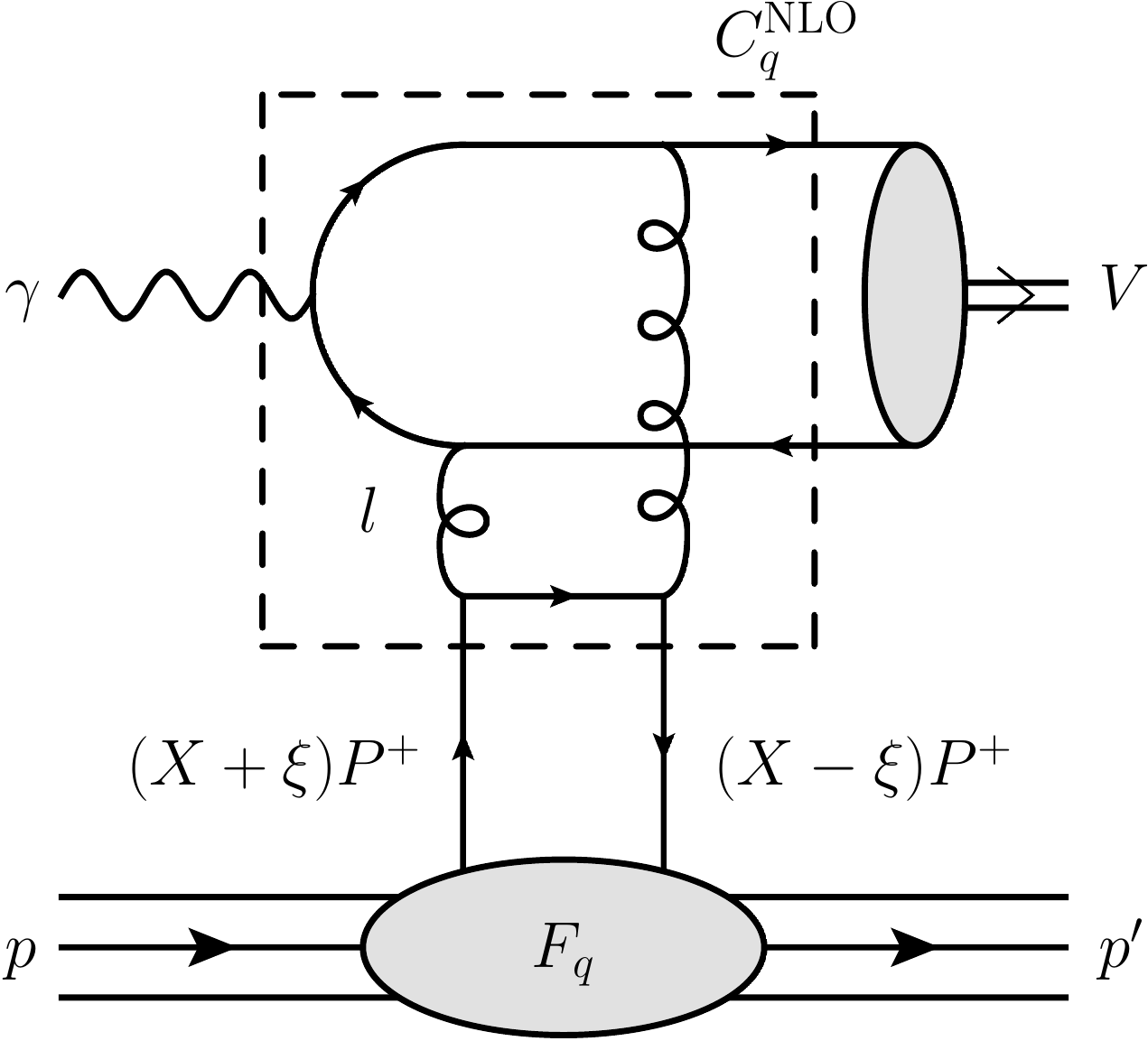}
    \caption{Five leg pomeron-like exchange diagrams at LO (left panel) and NLO (right panel). The perturbatively calculable coefficients are denoted by $C_{g,q}$ and the GPDs by $F_{g,q}$, with parton momentum fractions $x=X+\xi$ and $x' = X-\xi$. }
    \label{fig:my_label}
\end{figure}
In the case of photoproduction, the mass of the charm quarks, $m_c$, permits the use of pQCD and we describe the outgoing $J/\psi$ wavefunction within non-relativistic QCD (NRQCD), see also the study by Hoodbhoy \cite{Hood}.

What is probed in this process is the Generalised Parton Distribution Function (GPD), an off forward generalisation of the conventional collinear PDFs, in which we account for a skewing parameter, $\xi$, between the initial and final hadronic states, see \cite{Diehl} for a review. Fortunately, assuming that the input distribution has no singularities in the right-half of the Mellin-$N$ plane, one may relate the two outside the timelike region $X \in [-\xi,\xi]$ with $\mathcal O(x)$ accuracy  via the Shuvaev integral transform, see e.g. \cite{Cathy}. In the literature, analytic approximations to this poorly converging, computationally expensive transform in the so-called maximal skew regime, $X \sim \xi$, are often used but in this work we employ the full transform. The GPD grids are constructed from a three-dimensional parameter space in $X, \xi/X$ and scale $Q^2$ \cite{Cathy} with forward PDF grids taken from the LHAPDF \cite{LHAPDF} interface and suitably interpolated before being cast into the Shuvaev transform.  The grid is optimised such that we do not overly populate an area that results in a flat interpolation - having more points around $\xi/X \sim 1$, the border between the DGLAP and ERBL region, mitigates edge effects \cite{Cathy}.

\section{Stability of amplitudes at NLO and cross section predictions}
In the conventional approach, that is to say collinear factorisation within the $\overline{\text{MS}}$ scheme, the NLO contribution exhibits poor perturbative convergence (with the NLO correction greater or larger than LO and of opposite sign). Moreover, there exists a strong dependence on the factorisation scale $\mu_F$, as illustrated in Fig. 3. 
\begin{figure}[h]
    \centering
    \includegraphics[scale=0.45]{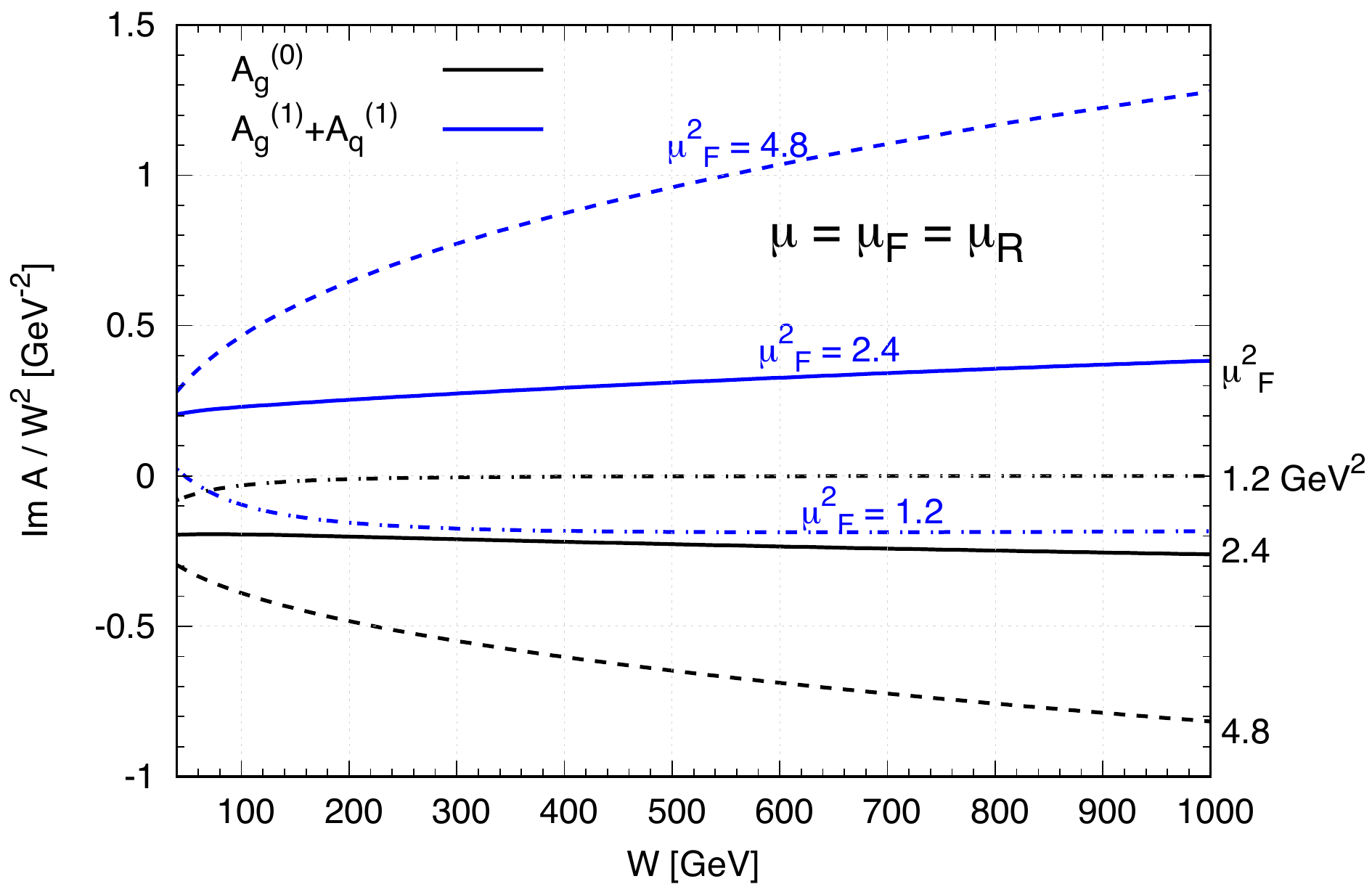}
    \caption{$\overline{\text{MS}}$ scale variations of $\text{Im}\, A/W^2$  at LO and NLO generated using CTEQ6.6 global partons at $\mu_F^2 = \mu_R^2 = 1.2,2.4,4.8\, \text{GeV}^2$. $\text{Im}\, A$ is the imaginary part of the amplitude. }
    \label{fig:my_label}
\end{figure}
A stepping stone towards improving this behaviour lies in the high energy asymptotics of the NLO contribution, which contains a double logarithm $\alpha_s \ln (1/\xi) \ln (\mu_F/m_c)$. In the small $x$ (small $\xi$) domain, this term leads to a large enhancement of the amplitude. However, we find that by choosing the factorisation scale\footnote{In our predictions, we take $\mu_F = m_c$.} $\mu_F \simeq m_c$ we minimise the double logarithmic contributions in the NLO contribution. 
This amounts to, via the setting of the factorisation scale, a shift of terms from the NLO contribution into the LO GPD. The remnant NLO coefficient function has a small, residual dependence on the factorisation scale, $\mu_f$, see \cite{dlog} for full details. Of course, it is possible to resum the BFKL-like contributions $\sim (\alpha_s \ln (1/\xi))^n$ which do not depend on the factorisation scale, see \cite{ivanovlog}. We do not do so here since, after their resummation, the coefficient function would mainly sample $x \sim \mathcal O(1)$ and we would lose the advantage of probing the unexplored small $x$ regime.
However, as shown in the left panel of Fig. 4, fixing the scale $\mu_F$ does not lead to a large reduction of the factorisation scale dependence of the amplitude.
Have we missed some effect? We now show the importance of a $Q_0$ cut in the computation of the coefficient function in order to avoid double counting.

The impact of a $Q_0$ cut was first demonstrated in \cite{Q0}. 
In starting the DGLAP evolution at a sufficiently perturbative scale $Q_0$, (i.e our input GPD is at scale $Q_0$) we are effectively double counting if we do not, in turn, subtract from the NLO coefficient function the region $|q|^2<Q_0^2$, where $q$ is the $t$-channel four momentum of the gluons. This region is systematically removed by restricting the virtuality of the loop momentum in the relevant ladder diagrams to be above $Q_0$, that is $|q|^2 > Q_0^2$. Such a $Q_0$ subtraction amounts to a power correction of $\mathcal O(Q_0^2/\mu_F^2)$ which for us is sizeable due to the factorisation scale being almost in the non-perturbative regime, $\mu_F = m_c$, so that the correction is $\mathcal O(1)$. The impact is clear: there is not much scope for evolution between the soft and (semi) hard sector whilst in, say, Higgs production we would have $\mu_F = \mathcal O(m_H)$ and the overall effect is negligible. As shown below in the right panel of Fig. 4, the inclusion of this cut results in a vastly improved theoretical prediction that is both stable and with LO and NLO contributions indicative of perturbative convergence.
\begin{figure}[h]
    \centering
    \includegraphics[scale=0.40]{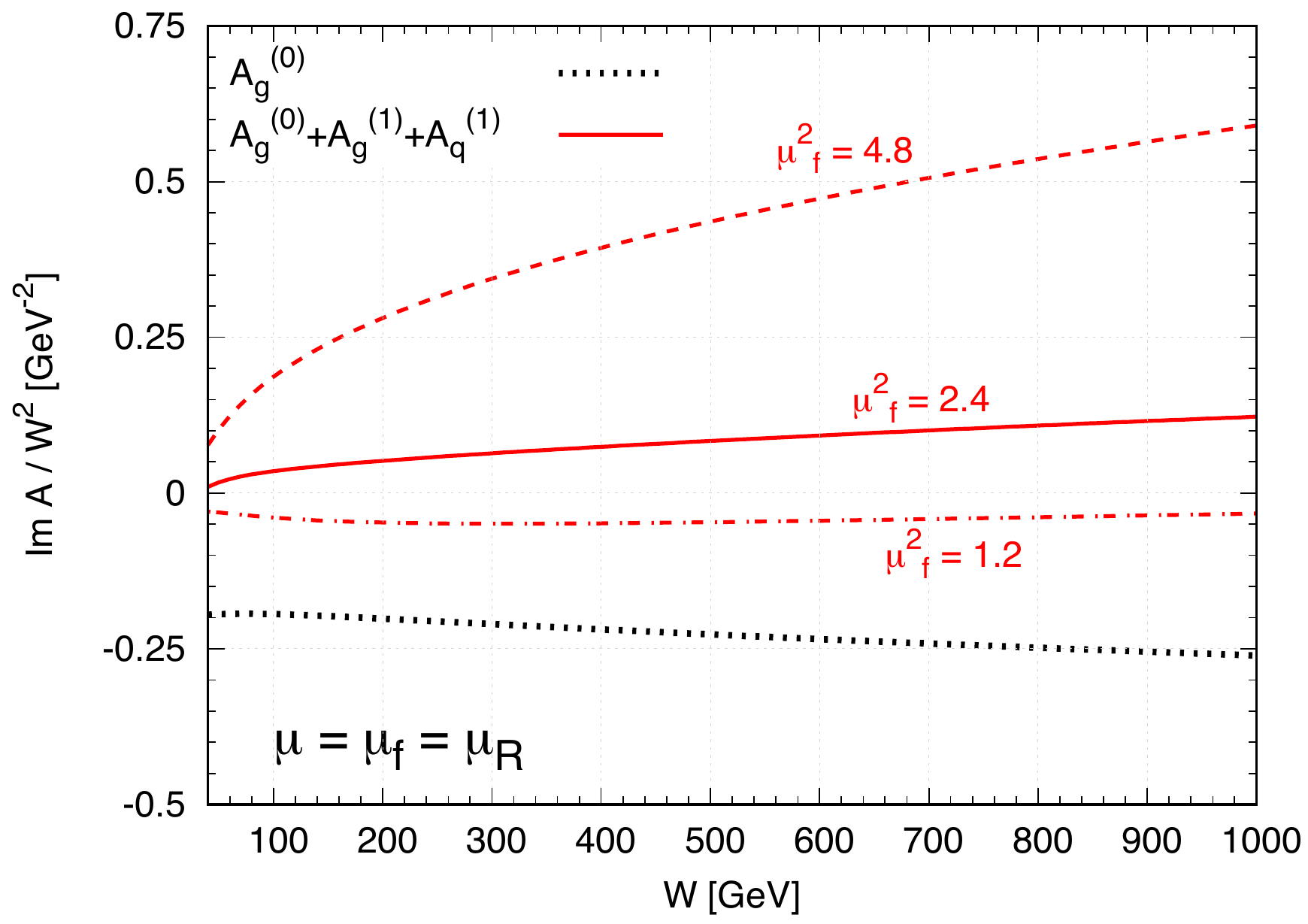}
    \quad
    \includegraphics[scale=0.40]{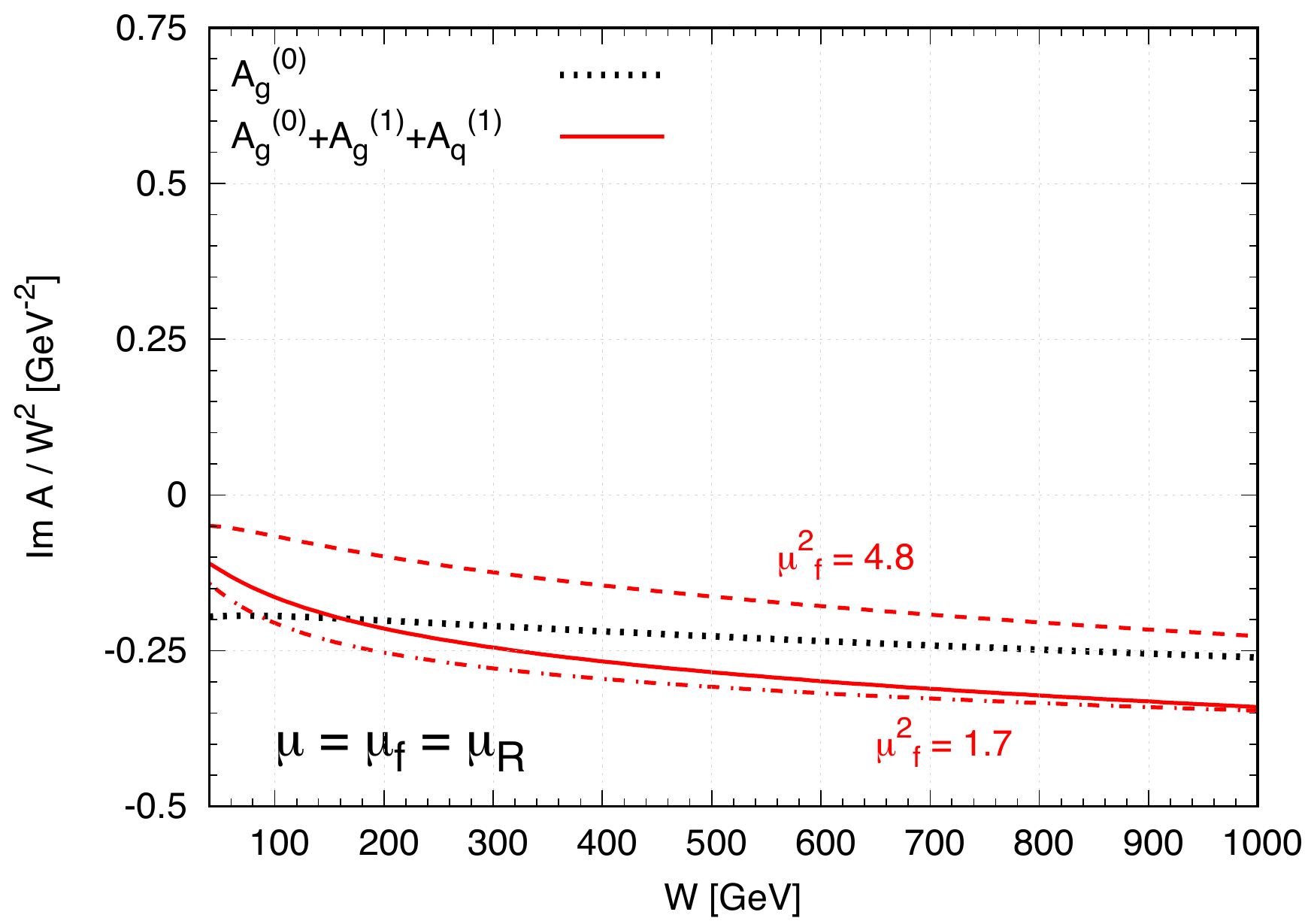}
    \caption{Results of $\text{Im}\, A/W^2$ \text{vs.} $W$ for the scale fixing procedure (left panel) and the implementation of a $Q_0$ cut (right panel), using CTEQ6.6 partons with $\mu^2_F = m_c^2 = 2.4\,\text{GeV}^2$ fixed.}
\end{figure}

With the NLO amplitudes sufficiently stable, let us now proceed to the cross section prediction. Recall that the DGLAP evolution starts at the scale at which we evaluate our input GPD, that is at $Q_0$. It ends at $\mu_F = m_c$, our `optimal' factorisation scale. Therefore, to maintain the typical hiearchy of scales, one would like $Q_0 \leq \mu_F$. Clearly, however, $Q_0$ should be of the order of the starting scale of the PDF fit.
We thus have only a small region of parameter space in which we can achieve stability. We also vary $\mu_f = \mu_R$ simultaneously, in line with the BLM scale prescription. Alleviation of the factorisation scale dependence upon imposition of the cut has paid dividends in leading to the dominance of the gluon contribution over the quark contribution, as we show below in the left panel of Fig. 5. In fact, after the $Q_0$ subtraction, the quark contribution is essentially zeroed so that the exclusive $J/\psi$ data is really a probe of the gluon density. Indeed, the data probe down to $x \sim 3 \times 10^{-6},$ where of course the gluon density is expected to be much larger than the quark density. 
\begin{figure}[h]
    \centering
    \includegraphics[scale=0.40]{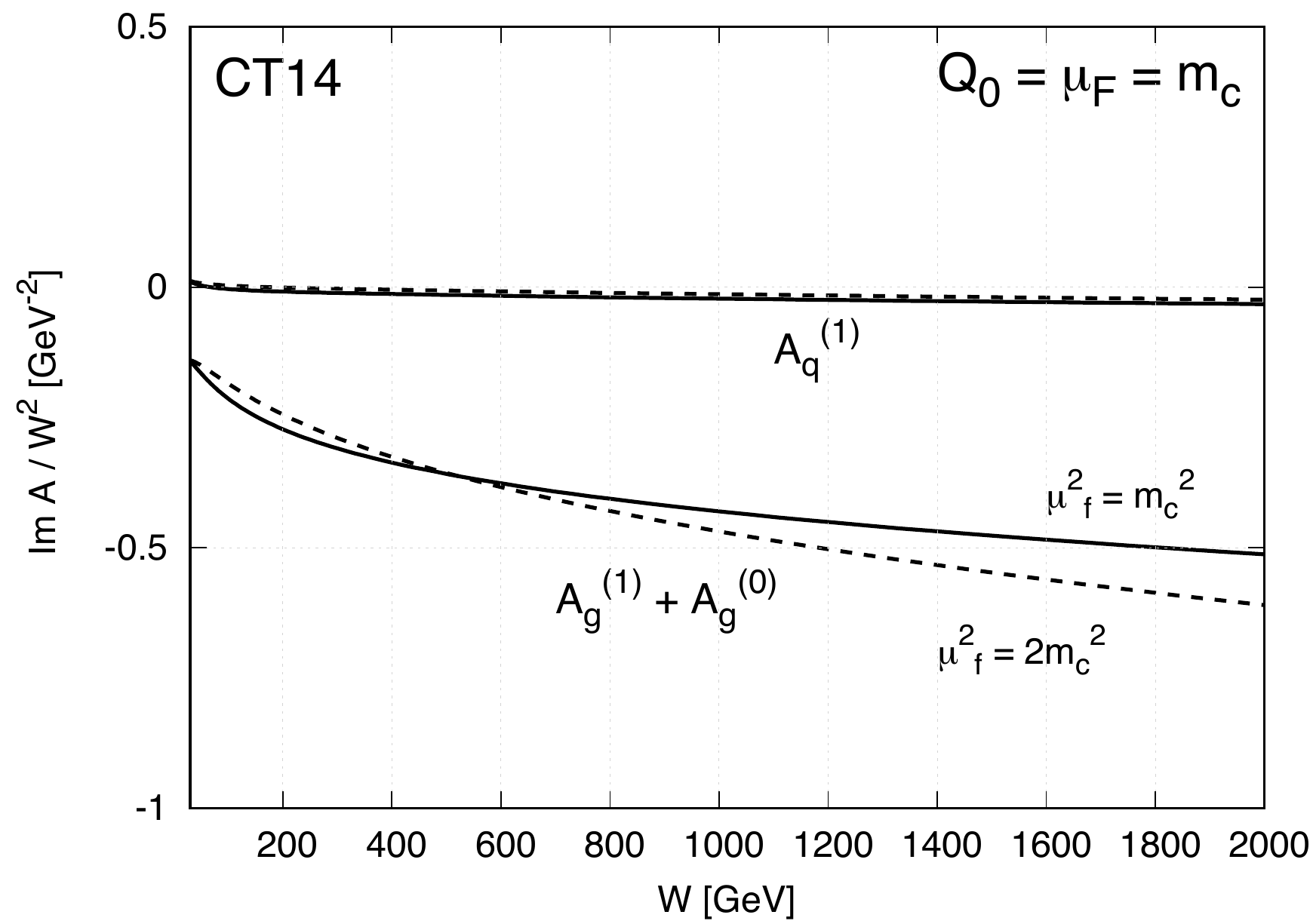}
    \quad
    \includegraphics[scale=0.43]{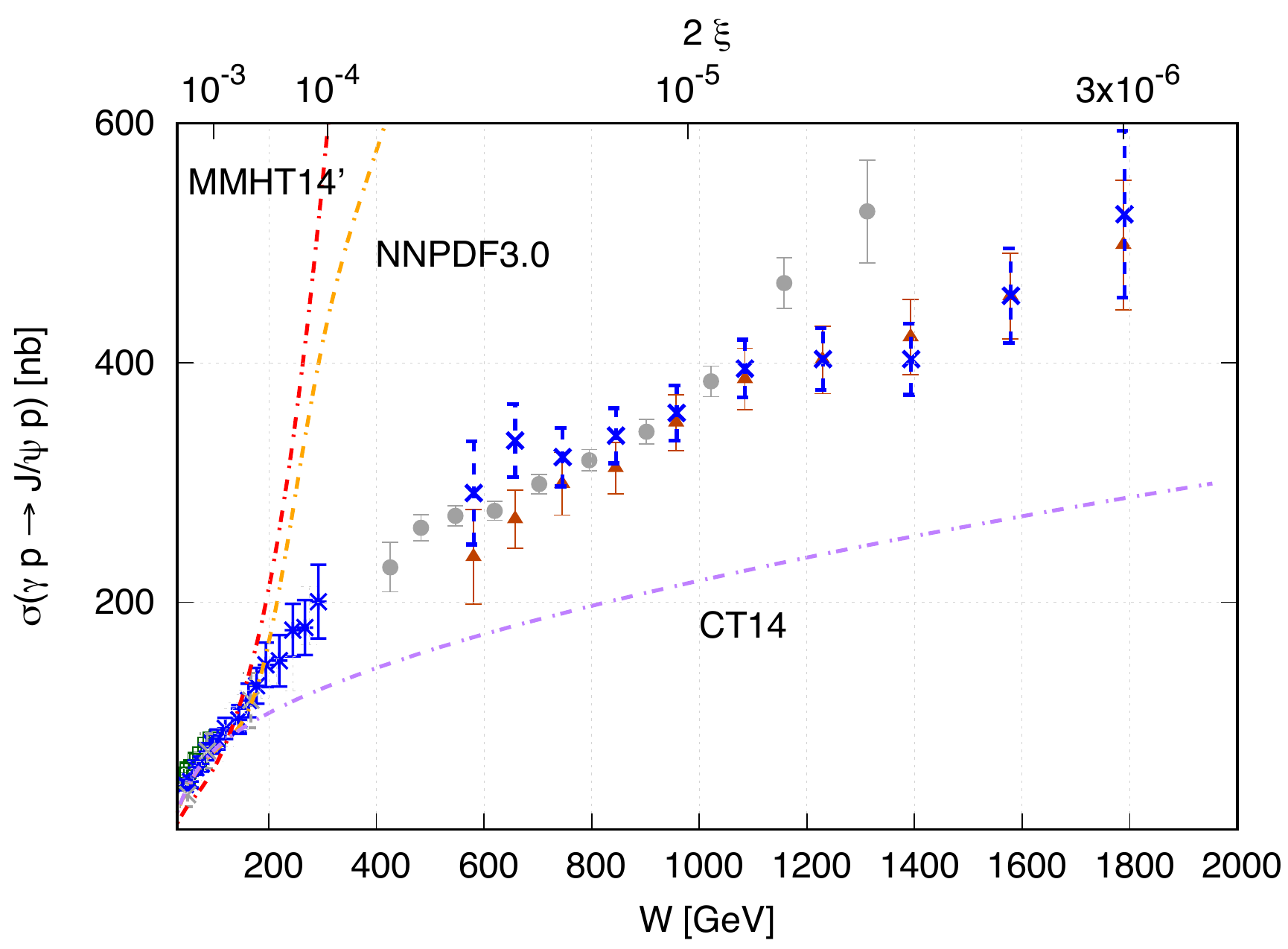}
    \caption{The quark contribution is seen to be almost vanishing and the scale dependence is small throughout the HERA and LHCb regions (left panel). Plot showing the large uncertainty between the global parton predictions in the LHCb regime. The typical $x \simeq 2 \xi$ values probed are shown on the upper axis. (right panel)}
\end{figure}

Within a given global fit, we obtain stability of the cross section prediction with $Q_0 = \mu_F = m_c$ and for scale variations $\mu^2_f = \mu^2_R \in [\,m^2_c,\, 2 m^2_c\,]$.
From Regge based arguments, the imaginary part of the amplitude is the dominant contribution, especially at high energy. However, we nonetheless proceed to incorporate the real part via the dispersion relation $$\frac{\text{Re} A}{\text{Im} A} \sim \frac{\pi}{2} \lambda(W) = \frac{\pi}{2} \frac{\partial \ln \text{Im} A/W^2}{ \partial \ln W^2}.$$

In Fig. 5, (right panel) we show cross section predictions (proportional to the square of the gluon PDF at $Q^2 \simeq m_c^2 = 2.4\,\text{GeV}^2$) using three sets of global PDF fits \cite{NNPDF3.0}, \cite{CT14}, \cite{MMHT14}, evaluated at $Q_0 = \mu_F = \mu_f = \mu_R = m_c$.  
The plot stresses the diversity of predictions we obtain in the LHCb regime\footnote{The huge error bands of the predictions, not shown, encompass the LHCb data.} whilst maintaining a degree of conformity in the HERA domain.
We see that the exclusive $J/\psi$ data will allow us to pin down the gluon PDF for $x \sim 10^{-5}$ for the first time. 
Of course, for the production of heavy final states at LHC energies, the PDFs are sampled at much higher factorisation scales $Q^2$ and momentum fractions $x$, and continue to be reliable. 
However, for low $x$ the gluon PDF serves as the boundary of BFKL-type physics giving information on confinement and saturation. 

\end{document}